\documentclass[10pt,dvipsnames]{article}
\usepackage[a4paper, margin=1in]{geometry}
\usepackage[utf8]{inputenc}
\usepackage[T1]{fontenc}
\usepackage{graphicx}
\usepackage{subfigure}
\usepackage{amsmath}
\usepackage{authblk}

\usepackage{tikz}
\usetikzlibrary{positioning}
\usepackage{braket}
\usepackage{bbold}
\usepackage{float}
\usepackage{times}
\usepackage{xcolor}
\definecolor{myurlcolor}{rgb}{0,0.3,0.6}
\definecolor{myrefcolor}{rgb}{0.8,0,0}
\usepackage[unicode=true,
bookmarks=false,bookmarksnumbered=false,
bookmarksopen=false, breaklinks=false,pdfborder={0 0 0},backref=false,
colorlinks=true, linkcolor=myrefcolor,citecolor=myurlcolor,urlcolor=myurlcolor]
{hyperref}

\definecolor{darkred}{rgb}{0.55, 0.0, 0.0}
\definecolor{greenish}{HTML}{228833}

\title{Quantum conference key agreement in asymmetric star networks with memory-multiplexing}
\title{Strategy optimization for quantum conference key agreement\\ in asymmetric star networks}
\author[1]{Janka Memmen}
\author[2]{Julia Kunzelmann}
\author[4]{Nathan Walk}
\author[4,5]{Jens Eisert}
\author[3,4]{Julius Wallnöfer}
\affil[1]{Electrical Engineering and Computer Science Department,
Technische Universit\"at Berlin, Einsteinufer 17, 10587 Berlin, Germany}
\affil[2]{Institute for Theoretical Physics III, Heinrich Heine University Düsseldorf, 40225 Düsseldorf, Germany}
\affil[3]{Institut f\"ur Theoretische Physik, Universit\"at Innsbruck, Technikerstra{\ss}e 21a, 6020 Innsbruck, Austria}
\affil[4]{Dahlem Center for Complex Quantum Systems, Freie Universit\"at Berlin, Arnimallee 14, 14195 Berlin, Germany}
\affil[5]{Helmholtz-Zentrum Berlin für Materialien und Energie, Hahn-Meitner-Platz 1, 14109 Berlin, Germany}
\date{}

\begin{document}

\maketitle

\begin{abstract}
    The distribution of entangled states is a core task for quantum networks facilitating quantum communication, and the use of multipartite entangled states comes with its own set of considerations. In this work, we analyze a quantum conference agreement protocol based on GHZ states in a network with a central station to which multiple clients are connected. Using comprehensive numerical simulations, we investigate how minor variations in the scenario—such as the number of parties, the number of memories, and asymmetric distances from the central station—can drastically influence the performance of the protocol. In particular, we demonstrate that it is crucial to adjust the strategy by optimizing cutoff times. From a broader perspective, we argue that numerical simulations are an indispensable tool for protocol design for devising realistic schemes for quantum communication.
\end{abstract}


\section{Introduction}
The field of quantum communication is concerned with making the various applications of quantum information theory, such as quantum key distribution \cite{cryptography_review2002, cryptography_review2018, cryptography_review2020} facilitating secret communication, available to distant parties. Research on addressing the challenges that arise specifically from the quantum nature of these systems has been so fruitful that even thinking of what is called the ``quantum internet'' \cite{Kimble2008, quantum_internet_wehner2018} has become reasonable, and quantum networks are now an established topic in the field. This is primarily due to the substantial experimental challenges associated with realizing the untrusted nodes required for quantum repeaters, but also to theoretical questions that arise when considering cryptographic applications of quantum communication \cite{cryptography_review2020,MarkWilde} beyond the generation of a secure key between two parties.

The distribution of entangled states is a key task that would be asked of such a quantum network. While quantum information theory has a wealth of protocols that revolve around the use of bipartite entangled states, there are a range of applications, for which multipartite entanglement are either required or provide an advantage, such as \emph{quantum conference key agreement} (CKA) and anonymous conference key agreement \cite{Murta_2020, Grasselli2022, Hahn2020, Webb2023}, quantum secret sharing \cite{Hillery1999} and quantum sensing \cite{quantum_sensing_review}. 
Distributing multipartite entangled states over long distances comes with its own set of considerations and, in general, the role multipartite entanglement should play in the operation of quantum networks is an issue that is not 
yet settled. While locally generating a multipartite state and using long-distance bipartite entanglement and quantum teleportation to distribute it to the appropriate parties is an option, it is not clear whether this is the most efficient way. In particular, multipartite entanglement \cite{HahnPappaEisert,Dahlberg} has been shown to be useful for speeding up quantum conference key agreement by overcoming bottlenecks in quantum networks \cite{Epping_2017}. Furthermore, conceptual directions for quantum networks such as entanglement-based quantum networks \cite{Pirker_2018, Pirker_2019} rely on a steady supply of distributed multipartite entangled states.

In this work, we focus on the following scenario: A star-shaped network topology with a central station that orchestrates the entanglement generation process between multiple parties. This type of setup is often called an entanglement switch, which has been studied before \cite{vardoyan2021, vardoyan2023, coopmans2022, netsquidmultipartite, netsquid}. Specifically, Ref.\ \cite{netsquidmultipartite} provides analytical formulas 
for the expected fidelity for the case of equidistant stations with one quantum memory per repeater link and no cutoff times and uses the quantum network simulator NetSquid \cite{netsquid} to verify the results numerically. In 
Ref.\ \cite{netsquid}, a similar setup has been studied, and it has been numerically assessed where either Bell pairs or 4-party GHZ states are distributed in a network with 9 parties and a central station. That setup uses multiple memories as a \textit{buffer} (which for finite buffer sizes has some similarities to a cutoff time). In Ref.\ \cite{memmen2023}, an advantage of multipartite entanglement has been demonstrated by deriving noise and distance thresholds for which the multipartite approach outperforms bipartite-entanglement–based CKA protocols in an asymmetric bottleneck scenario, where one party is significantly farther from the central station.

In this work, we opt to perform numerical simulations to capture the effects arising from combining multiple of these variations of the base scenario, as treating them analytically gets increasingly challenging as the protocols get more complex. The core challenge when analytically investigating these setups, especially asymmetric ones, is that generally one cannot simply analyze parts of the setup in isolation. Instead, one needs to consider how every decision in one part of the protocol affects all others (in particular due to the time-dependent nature of memory noise).

We demonstrate in multiple different scenarios that strategy optimization is not only a helpful extra but is actually essential in pushing the limits of what is possible with a given hardware setup. In particular, we show the importance of using a cutoff time strategy, i.e.,  discarding qubits that have been stored in memory after a certain time to lower the impact of memory noise. Specifically, optimizing the cutoff time allows one to obtain a non-zero key rate in parameter regimes where without it none would be possible. 

Furthermore, by investigating multiple similar scenarios, we show how subtle differences in the setup and requirements lead to vastly different outcomes. For example, the positioning of the entangled pair sources and whether the distributed states are to be used immediately or stored for later lead to tangible differences for the schemes considered here. Moreover, the protocols for these setups need to be optimized independently for each setup, e.g., the best choice of a cutoff time can be different. 

On a conceptual level, we emphasize the importance of systematically exploring such scenarios using numerical tools. This insight emerges not only from the present work, but also from a growing body of recent studies. The design of effective quantum communication protocols is inherently complex and multi-faceted, making it difficult to identify optimal operating regimes purely analytically. In this context, Monte Carlo simulations provide a powerful means of locating favorable parameter “sweet spots” and guiding protocol design. A central message of this work is therefore that simulation-driven exploration is not merely useful, but often essential, as purely analytical approaches are unlikely to suffice for identifying high-performance protocols.


The work is structured as follows. In Section \ref{sec:models} we describe in detail the scenarios and noise models we consider. In Section \ref{sec:results_symmetric} we present our simulation results for symmetrically positioned stations and discuss the impact of having multiple memories, a variable number of parties and optimizing cutoff times. Then, in Section \ref{sec:results_asymmetric} we show how having a bottleneck network, where one station is much farther away from the central station than others, impacts the key distribution rate and how the use of cutoff times is particularly valuable in this setup. There, we also demonstrate that our simulation is able to handle all the aspects mentioned above at the same time in a network between multiple German universities. Finally, in Section \ref{sec:summary} we summarize our findings and discuss potential further research directions.

\section{Models and protocols}
\label{sec:models}

We begin by introducing the basic setup and premise of this work and establishing the notation.
We consider the following setup: A central station $C$ is connected by quantum channels to each of $N$ clients $\{B_1,\dots, B_N\}$. The central station has $m$ quantum memory slots per client that are allocated to that specific link and, depending on the protocol, the clients may also control $m$ quantum memory slots each.
Using entangled pair sources, the clients establish entangled pairs with the central station. Once a pair is present for each client, a joint measurement on the qubits at the central station is performed to join the states together to an $N$-party GHZ state shared by the clients.

\subsection{Noise models}
Noise and imperfections are core factors that need to be considered for entanglement distribution protocols. After all, the core point for thinking about quantum repeaters is to combat errors and losses along the way of entanglement generation.

\paragraph{Arrival probability.}
A principal factor in distributing entangled states is that it likely has to be attempted multiple times. This is due to loss, i.e.,  a sent qubit may not arrive at its destination. We model the chance that a qubit successfully passes through a quantum channel as $\eta_{\text{ch}} = e^{-l/L_{\text{att}}}$, where where $l$ is the length of the channel and $L_{\text{att}}=22 \text{\ km}$ the attenuation length in an optical fiber at telecom wavelengths.
Furthermore, there are other per-link overheads that are probabilistic and therefore may cause an unsuccessful attempt,, e.g., wave length conversion efficiency, detector efficiency or memory efficiency, which we summarize as one parameter $P_\text{link}$. The overall efficiency is  then given by $\eta = P_\text{link} \cdot \eta_\text{ch}$.

\paragraph{Dark counts.}
Additionally, there is the chance that due to dark counts a state is accepted even though no qubit is there. So, the effective chance that we think an entangled state has been established is given by $\eta_\text{eff} = 1 - (1-\eta)(1-P_D)^2$ where $P_D$ is the chance that a dark count happens in the time interval where a click would be expected. 
We model mistakenly accepting a state by replacing the qubit with a fully mixed state as
\begin{equation}
\mathcal{D}^{(i)} \rho = \alpha(\eta) \rho + [1 - \alpha(\eta) ] \; \mathrm{tr}_i(\rho) \otimes \frac{\mathbb{1}^{(i)}}{2} \quad \text{with} \quad\alpha(\eta) = 
\frac{\eta (1-P_D)}{\eta_\text{eff}}.
\end{equation}
In particular, this means that the effect of dark counts is inherently tied to the probability of a true click to occur, i.e.,  if $\eta$ is very low, nearly every click is a dark count. 
Here, we consider $P_D = 10^{-6}$, which, e.g., would be consistent with a dark count rate of 1~Hz and a $1\text{\ } \mu\mathrm{s}$ detection window.

\paragraph{Initial noise.} In addition to the noise that arises from dark counts, the preparation of the initial entangled states may be imperfect or there are other systematic per-link errors, which we model with an initial fidelity $F_\text{init}$ as
\begin{equation}
    F_{\text{init}} \ket{\Phi^+}\bra{\Phi^+} + \frac{1 - F_{\text{init}}}{3} (\ket{\Phi^-}\bra{\Phi^-} + \ket{\Psi^+}\bra{\Psi^+} + \ket{\Psi^-}\bra{\Psi^-})
\end{equation}
Throughout this work, we will consider $F_\text{init}=0.99$. {Fidelities greater than this have been measured with photonic Bell states at telecoms wavelengths \cite{psiquantum_exp}.}

\paragraph{Memory noise.}
An essential part of advanced protocols is the access to quantum memories. Qubits stored in quantum memories are affected by a time-dependent noise, which we assume is biased towards one basis, hence we model it as dephasing noise according to
\begin{align}
    \mathcal{E}_z^{(i)} (t) \rho = (1 - \lambda (t)) \rho + \lambda (t) Z^{(i)} \rho Z^{(i)}
    \end{align}
    with 
    \begin{align}
    \lambda (t) = \frac{1 - e^{-t/T_\text{dp}}}{2}
\end{align}
where $T_\text{dp}>0$ 
is the dephasing time, a parameter describing the quality of the quantum memory.    

\subsection{Scenarios and protocols}
In this work we consider two different goals: 
\begin{itemize}
\item The distribution of a multipartite entangled GHZ state shared by all participating clients $B_i$, which may later be used by the clients for any application. In this case the outer stations need to have a quantum memory in which their qubits are stored. We will refer to this scenario as \emph{Distribute} (Dis). 
\item Immediately making use of the connection for a quantum conference key protocol. In this case, memories are required by the server but not by the clients. Clients can instead immediately measure their qubits and figure out which of their measurement results to use in post-processing. We will refer to this scenario as \emph{Measure} (Meas).
\end{itemize}
While we use the key rate for quantum conference key agreement to assess both of these scenarios (as it captures both the rate and the quality of the distribution process), it should be noted that there is a fundamental difference between actually having a target state physically present at some point in time and just making use of the quantum correlations, which has implications for the network design and capability of the network nodes.

\subsubsection{Distribution times}
Since loss plays a significant role in the distribution of the initial pairs, many trials are necessary until one is successful. The number of trials $k$ until the next successful trial is geometrically distributed with a success probability of $\eta_\text{eff}$. 
It takes a preparation time $T_P$ to generate an entangled pair, which can then be sent through the channel.
When qubits need to be stored in memory, one needs to wait for confirmation of the other side of the link on whether the attempt was successful, which takes $2l/c$ where $c$ is the communication speed (we take $c=2 \times 10^8 \text{ m/s}$, which is the speed of light in optical fiber at telecom wavelengths). Therefore, the time for one trial is given by $t_\text{trial} = T_P + 2l/c$ if the generating party needs to wait for a confirmation. These trial times are the elementary time-steps for the individual links to establish a Bell pair, which can be attempted independently in parallel for each client.
During the time it takes for the confirmation to arrive, the qubits in memory have already been affected by the memory noise for $t_C$ or $t_B$ at the central station or the clients, respectively. The details vary depending on where the entangled pair sources are located. 

Furthermore, if the source is positioned at the outer station in Scenario Measure, the half of the pair at the client can be measured immediately and the next attempt started immediately, as the client does not need the information on whether the trial is successful and the central station can be assured that a qubit that actually arrives can be immediately used. As such, assuming this process is done continuously and the central station has some kind of mechanism to block excessive arriving qubits, the trial time is simply given by the generation rate of the source, i.e., $t_\text{trial} = T_P$.
These higher rates depending on the position of the sources is something that is discussed for bipartite repeaters; see, e.g., Refs. \cite{qlinkx_whitepaper, piparo2017}, which also emphasize that the availability of fast sources is dependent on the physical platform. However, it should be noted this difference vanishes if one considers more than two repeater links.
In Table \ref{tab:distribution_times} the distribution times and the accumulated memory noise until the state is ready to be used by the central station are summarized for all scenarios.

\begin{table}[htbp]
    \centering
    {\renewcommand{\arraystretch}{1.3}
    \begin{tabular}{cc|cccc}
         Scenario & Source location & $t_\mathrm{trial}$ & $t_C$ & $t_B$ & $t_s$\\
         \hline
         Distribute& $C$ &  $T_{\text{p}} + 2\frac{l}{c}$ & $2\frac{l}{c}$ & $\frac{l}{c}$  & $k\cdot t_{\text{trial}}$ \\
         Distribute& $B$ & $T_{\text{p}} + 2\frac{l}{c}$ & $0$ & $\frac{l}{c}$  & $k\cdot t_{\text{trial}} - \frac{l}{c}$\\
         Measure& $C$ & $T_{\text{p}} + 2\frac{l}{c}$ & $2\frac{l}{c}$ & $0$ & $k \cdot t_{\text{trial}}$ \\
         Measure & $B$ & $T_{\text{p}}$ & $0$ & $0$ & $k \cdot t_{\text{trial}}$
    \end{tabular}
    }
    \caption{With loss distributing entangled states takes multiple trials. Summary of timing information depending on the considered scenarios:  The time $t_s$ until the next successful trial happens and the central station is allowed to act on the corresponding qubit depends on the trial time $t_\text{trial}$ for one trial, the preparation time $T_\text{p}$, the distance $l$ is the distance between the outer station and the central station and the communication speed $c$.  $k$ is drawn from a geometric distribution with the success probability of one trial $\eta_\mathrm{eff}$.}
    \label{tab:distribution_times}
\end{table}

\subsubsection{Cutoff times}
One very useful addition to basic protocols is the addition of a so-called cutoff time strategy. If a qubit waits in memory for too long, the state will have decohered beyond being useful, potentially even becoming completely disentangled. As such, it can be beneficial to simply discard states that have waited longer than a chosen cutoff time $t_\text{\text{cut}}$ (see, e.g., Ref.\ \cite{Rozpedek_2018}). Note that this is an active protocol decision to discard the qubits and therefore is different from the maximum memory life times that are inherent to some quantum memories. Essentially, this strategy introduces a trade-off between the raw rate and the fidelity of the distributed GHZ states (or the quality of correlations in the CKA protocol) and as such needs to be optimized for the desired figure of merit. 

\subsubsection{Memory multiplexing}
When multiple memories are available per link, there is an additional choice to be made in the protocol, namely which Bell pairs to use if multiple states have been established at the same link. For our protocols, we chose to always use the most recently established pair for each link. For the star network, aspects of this have been discussed in more detail 
in Ref.\ \cite{Kunzelmann_2024}.
This means that we implicitly assume that each memory is individually addressable and can be jointly acted upon for any combination of memory slots.

\subsection{Conference key agreement rate}
\emph{Conference key agreement} (CKA) \cite{Murta_2020} is a natural extension of \emph{quantum key distribution} (QKD) to a multipartite setting, where multiple parties aim to establish a shared, so-called conference key. A multipartite entangled resource for this is the \emph{Greenberger–Horne–Zeilinger} (GHZ) state. For $N$ parties ($B_1,\ldots,B_N$) the GHZ state vector reads
\begin{align}
\ket{\text{GHZ}} = \frac{1}{\sqrt{2}} \left( \ket{0,\ldots,0} + \ket{1,\ldots,1} \right)_{B_1 ,\ldots , B_N}.
\label{GHZ_state}
\end{align}
In this work, we will focus on the $N$-party generalization of the BB84 protocol for CKA ($N$-BB84), where all parties switch between two different measurement bases: the $X$ and $Z$ bases. Given the nature of the GHZ state, rounds in which all parties measure in the $Z$ basis can be used for key generation, while rounds in which all parties measure in the $X$ basis can be used for parameter estimation.

In the limit of infinitely many exchanged bits, the secure key rate for the $N$-BB84 protocol is given by~\cite{Grasselli_2018, Proietti_2021}
\begin{align}
K_{\mathrm{CKA}} = Y \left[1 - h_2(Q_X) - \max_{2 \leq i \leq N} h_2(Q_{B_1, B_i})\right],
\label{CKA_rate}
\end{align}
where $Y$ denotes the overall yield, $Q_X$ the 
\emph{quantum bit error rate} (QBER) observed in the $X$ basis, $Q_{B_1, B_i}$ the QBER between the dealer $B_1$ and the $i$-th Bob in the $Z$ basis, and $h_2(\cdot)$ the binary Shannon entropy. Note that $Q_X$ is based on the collective $X$-measurement, while $Q_{B_1, B_i}$ is an individual QBER between the dealer $B_1$ and the $i$-th Bob. 

\paragraph{Yield.}
The yield represents the fraction of distributed states that successfully arrive at their intended destinations, enabling the execution of the protocol. It depends on the link success probabilities and the network architecture, and it essentially determines the rate at which the protocol can operate. Throughout this work, we express the yield per unit time. In contrast to yield per channel use, this corresponds to the raw rate at which multipartite entangled states can be generated. Although in simpler scenarios the yield could be calculated analytically, we will, throughout this work, always obtain it from simulation results for all scenarios considered.

\paragraph{QBER calculation.}
We will denote the noisy output state obtained from the simulation as $\rho_{B_1, \ldots , B_N}$. $Q_X$, the probability that a collective $X$-measurement yields an outcome that is discordant with the corresponding outcome of the noiseless state given in Eq.~\eqref{GHZ_state}, is expressed as
\begin{align}
Q_X = \frac{1 - \langle X^{\otimes N} \rangle}{2},
\end{align}
where $\langle X^{\otimes N} \rangle = \mathrm{Tr}\left(\rho_{B_1, \ldots , B_N} X^{\otimes N}\right)$. Note here that this is the QBER that will be influenced by the memory dephasing.
Analogously, the bipartite $Z$-basis QBER between the dealer (located at position $1$) and the $i$-th party is given by
\begin{align}
Q_{B_1, B_i} = \frac{1 - \langle Z_1 Z_i \rangle}{2}.
\end{align}
Contrary to $Q_X$, $Q_{B_1, B_i}$ is a two-qubit QBER and is only affected by noise of two qubits. 



\section{Simulation results}
\label{sec:results}

We now turn to discussing our results from comprehensive simulations.
Unless otherwise stated, all simulation results are obtained based on $10^{5}$ established GHZ states, or GHZ like correlations. 

\subsection{Symmetric star network}
\label{sec:results_symmetric}

\begin{figure}[h]
    \centering
    
\begin{tikzpicture}[
    scale=0.75, transform shape,
    node distance=2.5cm, 
    every node/.style={font=\sffamily},
    circleNode/.style={circle, draw, minimum size=0.8cm, align=center, fill=cyan!20!white}
]
  \node[draw, minimum size=2.7cm, shape=rectangle, rounded corners=8pt, thick] (center) {$C$};

  \node[circleNode] (inN) at (0,0.8) {QM}; 
  \node[circleNode] (inE) at (0.8,0) {QM}; 
  \node[circleNode] (inS) at (0,-0.8) {QM}; 
  \node[circleNode] (inW) at (-0.8,0) {QM}; 

  \node[circleNode] (N) at (0,4) {$B_1$};
  \node[circleNode] (E) at (4,0) {$B_2$};
  \node[circleNode] (S) at (0,-4) {$B_3$};
  \node[circleNode] (W) at (-4,0) {$B_N$};

  \draw[thick] (N) -- (inN) node[midway, right] {$l$};
  \draw[thick] (E) -- (inE) node[midway, above] {$l$};
  \draw[thick] (S) -- (inS) node[midway, right] {$l$};
  \draw[thick] (W) -- (inW) node[midway, above] {$l$};
  
  \draw[dotted, thick]
  ([shift={(-1.5,0.5)}]S.west)
    to[bend left=15]
  ([shift={(0.5,-1.5)}]W.south);

\end{tikzpicture}
\caption{Star network representing a symmetric single-memory multipartite quantum repeater. The $N$ parties $B_i$ are located around the central station $C$ at equal distances $l$. The central quantum repeater provides a single \emph{quantum memory} (QM) for each party. }
    \label{fig:star-network-topology}
\end{figure}
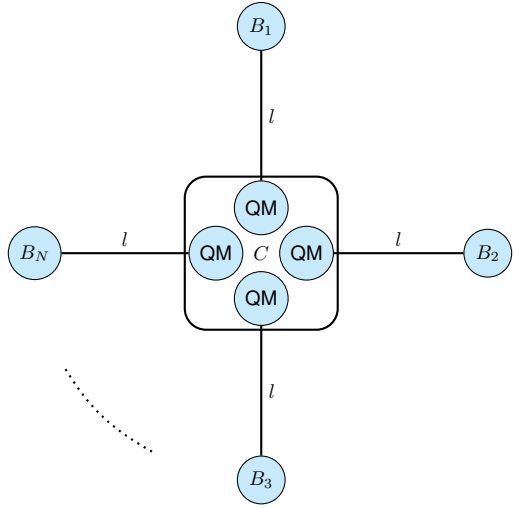

\subsubsection{Single memory per client}
For each scenario, i.e. the two strategies, \emph{Distribute} and \emph{Measure}, each paired with the two possible source locations B and C, we evaluate both the achievable fidelities and the asymptotic key rates (see Equation~\ref{CKA_rate}). This analysis is done for the case when the memories do not have a cutoff time and once with an applied cutoff time of $t_{\text{\text{cut}}} = 0.3\ \mathrm{s}$. 

In the case without a cutoff time, we find that the Meas-B scenario, where the sources are located at the clients, yields by far the best results, both in terms of achievable state quality and asymptotic key rate. However, this configuration is also the most challenging to realize experimentally, as it would require a mechanism to block incoming photons from entering the memory when it is already occupied. For this scenario and the chosen parameter set, even without a cutoff time, a single-link distance of slightly more than $200\ \mathrm{km}$ can be spanned, corresponding to an overall distance of more than $400\ \mathrm{km}$. Also, there is essentially no difference between the results with and without a cutoff time, as the applied cutoff time is simply too large to show any effect and actually discard qubits. Note here that this will be different when each link consists of repeater stations. For all other scenarios, however, the situation is markedly different. In particular, for both Dis scenarios and Meas-C, the introduction of a cutoff time enhances the fidelity of the multipartite entangled target state. It is to note that the two Dis configurations, where the sources are located either at the Bobs' stations or at the central station, show very similar behavior. 

Without cutoff times, the fidelity of the target state decreases significantly for channel lengths in the range of $50\ \mathrm{km} \leq l \leq 150\ \mathrm{km}$. When cutoff times are introduced, this decrease can be partially compensated for distances between $100\ \mathrm{km} \leq l \leq 165\ \mathrm{km}$. However, at these longer distances the fidelity is already too low to distill any secret key for a CKA protocol, as the corresponding QBERs are too close to the QBER threshold.

\begin{figure}[H]
    \centering
    \begin{minipage}{0.48\textwidth}
        \centering
        \includegraphics[width=\linewidth]{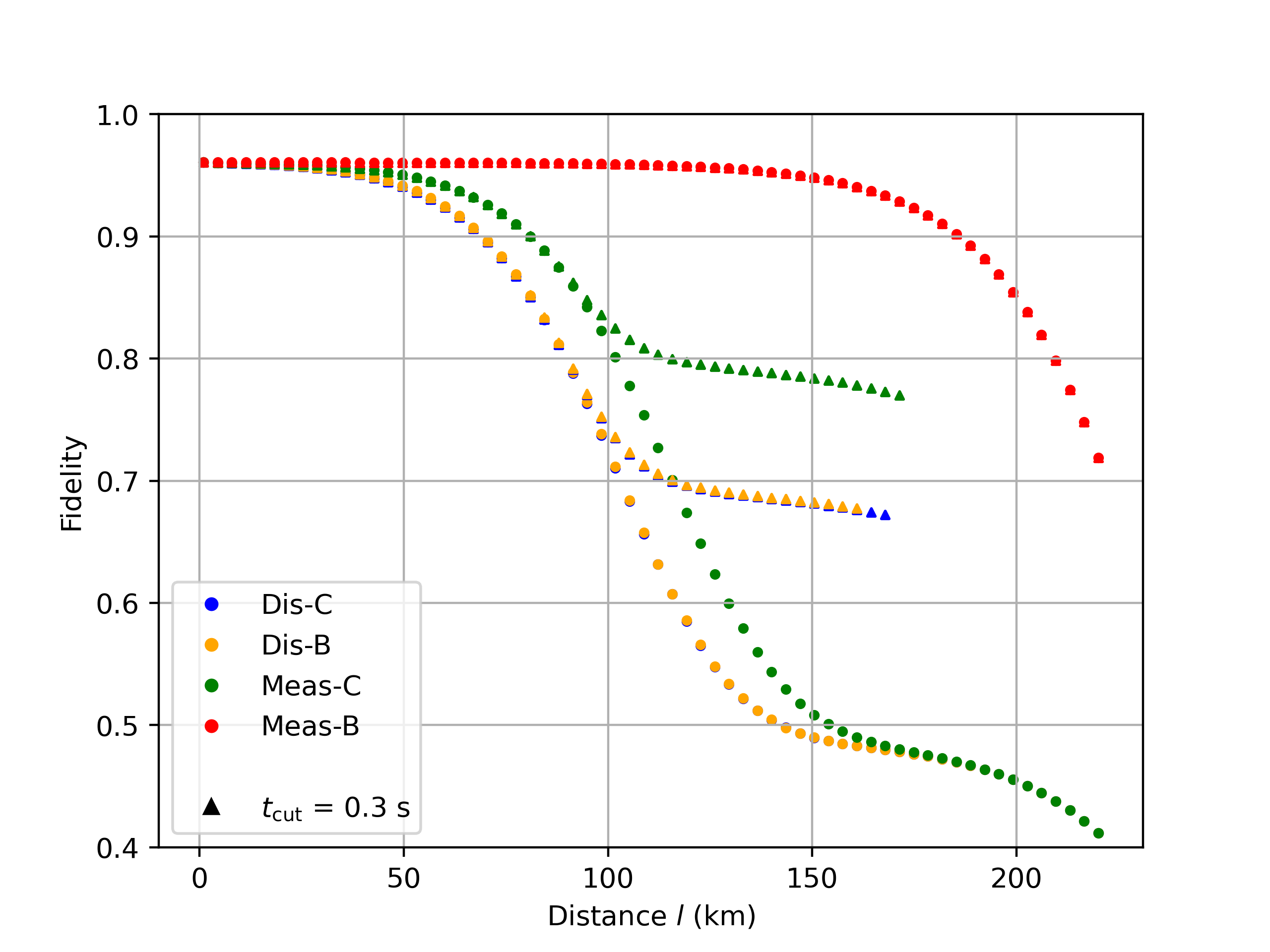}
        
        \small (a)
        \label{fig:sym_fid_nocut}
    \end{minipage}
    \hfill
    \begin{minipage}{0.48\textwidth}
        \centering
        \includegraphics[width=\linewidth]{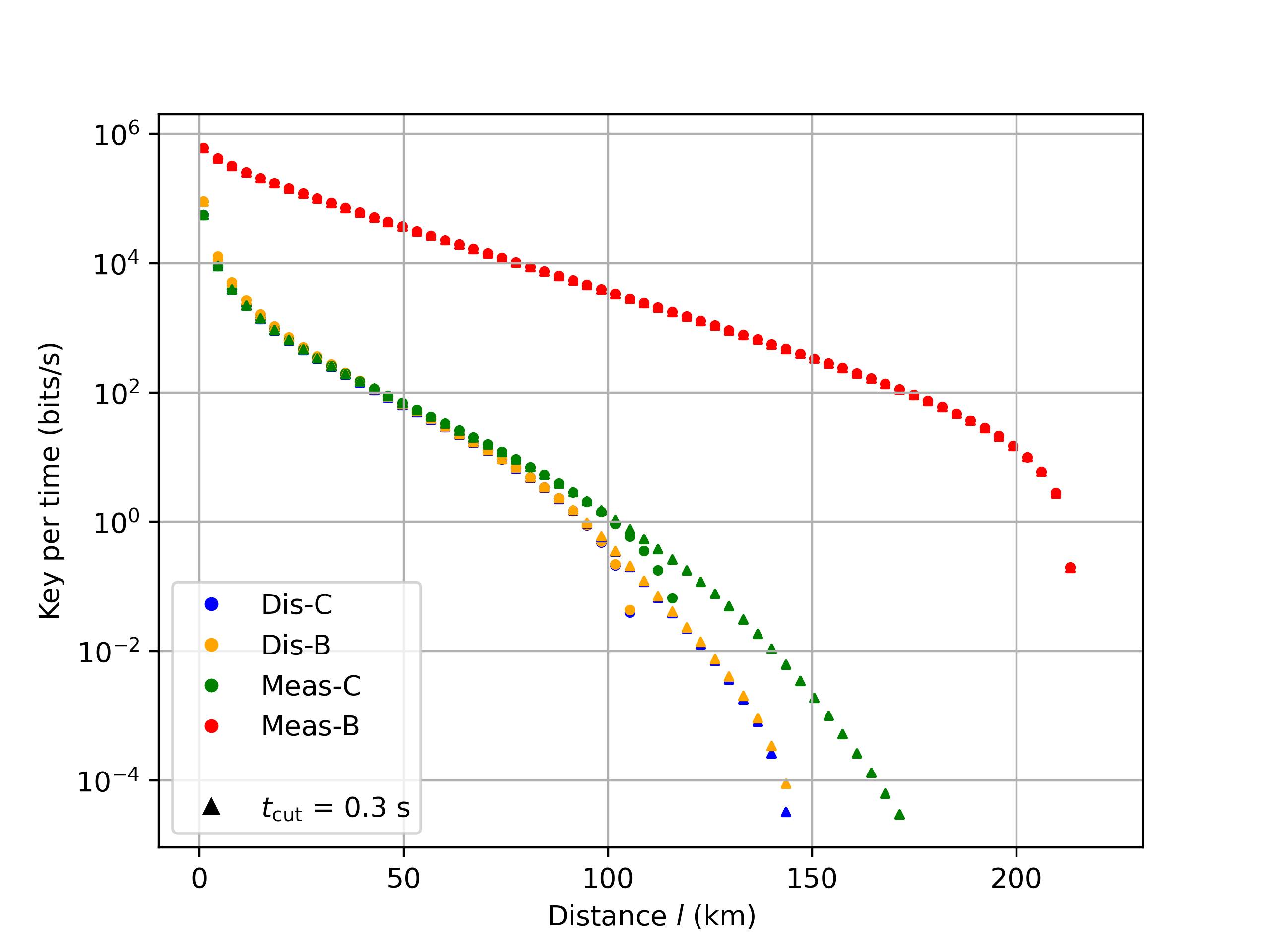}
        
        \small (b)
        \label{fig:sym_key_nocut}
    \end{minipage}
    \caption{Fidelities (a) and key rates (b) for the different strategies in dependence of a single link's distance for a fully symmetric star network with $N=4$ participants. The different markers symbolize whether a cutoff time of $t_\text{\text{cut}}=0.3\ \mathrm{s}$ was used or not. Other parameters are chosen as: $P_{\text{LINK}} = 1$, $F_{\text{INIT}} = 0.99$, $T_P = 10^{-6}\ \mathrm{s}$, $P_D = 10^{-6}$, $T_\text{dp} = 1\ \mathrm{s}$, $m= 1$.}
    \label{fig:sym_mem1}
\end{figure}

In contrast, for the Meas-C scenario, the fidelity is higher when the effect of the cutoff time becomes visible, allowing secure key generation to persist. The key rate depends on both the quality of the generated target state and the generation rate. While the application of a cutoff time improves the fidelity of the target state, it  reduces the generation rate, as some states are discarded and must be regenerated. Both of these effects can be seen in Figure~\ref{fig:sym_key_nocut}: the cutoff time can enable key generation beyond the previous threshold distance of $l \approx 110\ \mathrm{km}$ for Meas-C. However, the key rate constantly decreases and ultimately vanishes altogether for $l \geq 150\ \mathrm{km}$.

\subsubsection{Memory multiplexing}
\begin{figure}[H]
    \centering
    \begin{minipage}{0.48\textwidth}
        \centering
        \includegraphics[width=\linewidth]{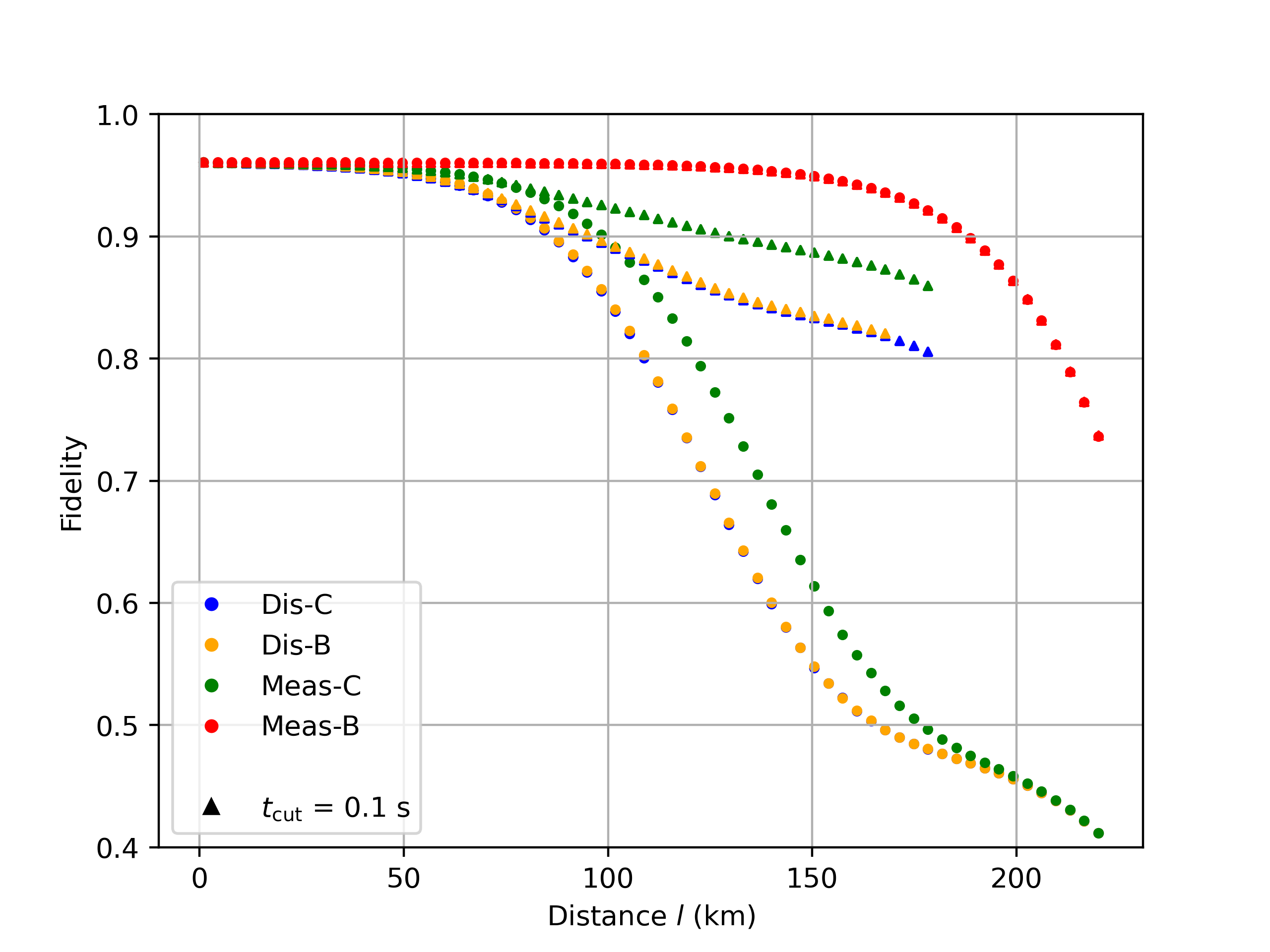}
        
        \small (a)
        \label{fig:sym_fid_5mem}
    \end{minipage}
    \hfill
    \begin{minipage}{0.48\textwidth}
        \centering
        \includegraphics[width=\linewidth]{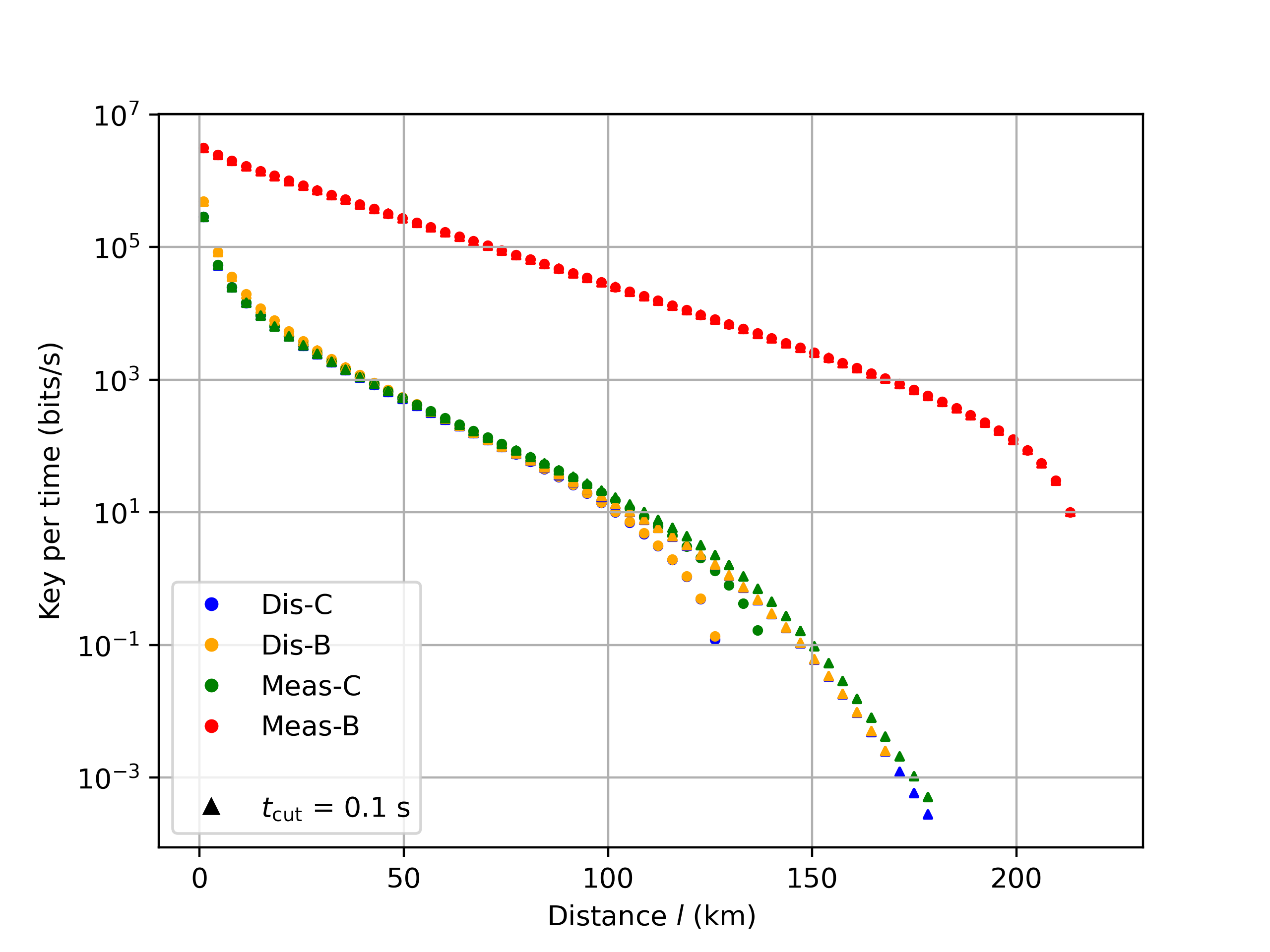}
        
        \small (b)
        \label{fig:sym_key_5mem}
    \end{minipage}
    \caption{Fidelities (a) and key rates (b) for the different strategies in dependence of a single link's distance $l$ for a fully symmetric star network with $N=4$ parties, when every party can make use of $m=5$ quantum memories. The different markers symbolize whether a cutoff time of $t_\text{\text{cut}}=0.1\ \mathrm{s}$ has been used or not. Other parameters are chosen as $P_{\text{LINK}} = 1$, $F_{\text{INIT}} = 0.99$, $T_P = 10^{-6}\ \mathrm{s}$, $P_D = 10^{-6}$, $T_\text{dp} = 1\ \mathrm{s}$.}
    \label{fig:sym_mem5}
\end{figure}

We now introduce multiple memories per party (the number of memories is denoted as $m$) to store arriving qubits at the center $C$. Consequently, up to $m$ entangled states can be performed in parallel for every attempt. Qubits that are not used in one attempt remain in the memory for the next one. To reduce storage times in noisy memories and therefore decoherence, we always choose the newest qubits first for the entanglement distribution, following  Ref.\ \cite{Kunzelmann_2025a}. Note that we consider symmetric setups here, i.e., each party is located at the same distance around the center.
Analogously to the single memory case, we again show the improvement in terms of fidelity (Figure~\ref{fig:sym_fid_5mem}) and secret key rate (Figure~\ref{fig:sym_key_5mem}) with and without cutoff times included. Note here that the cutoff time is set lower for the multi memory case (i.e., $t_{\text{cut}} = 0.1\ \mathrm{s}$), as the qubits generally spend less time in each memory, as multiple memories are available.

Similarly to the previous results, the Meas-B scenario, in which the source is located on the clients side, results in the best fidelities and key rates. The introduction of cutoff times does not influence the results. In contrast, the cutoff time plays an essential role in all other cases. The fidelity can be significantly increased for distances $l \geq 100\ \mathrm{km}$. The same holds for the key rate. An increase is achieved for both Dis scenarios and the Meas-C scenario. In all cases, a positive key rate for distances up to $l = 170\ \mathrm{km}$ can be reached. In contrast to the single memory setup, the Dis scenarios also reach these distances, such that the Meas-C scenario performs only slightly better. 
In total, higher key rates can be achieved by introducing multiple memories, because it takes less time on average to fill at least one memory per party. Consequently, the storage times decrease, and thus the fidelities increase (compare Figure~\ref{fig:sym_fid_5mem} taking into account the cutoff times). Accordingly, the secret key rate increases, yielding significantly higher rates compared to the single memory scenario. In particular, the improvement exceeds what would be achieved by simply repeating the protocol independently $m$ times in parallel.

\subsubsection{Optimization of cutoff times}
As seen in the previous sections, cutoff times can yield a significant advantage in both the maximally achievable communication distance and the fidelity of the generated states. However, they must be chosen carefully: If the cutoff time is too short, it becomes impossible to generate any multipartite entangled states. If it is too long, it has no effect since qubits are never discarded. This trade-off is a well-known result for repeater-chain architectures~\cite{requsim} (see also Ref.\ \cite{Satellite}).

\begin{figure}[h] 
    \centering
    \includegraphics[width=0.67\textwidth]{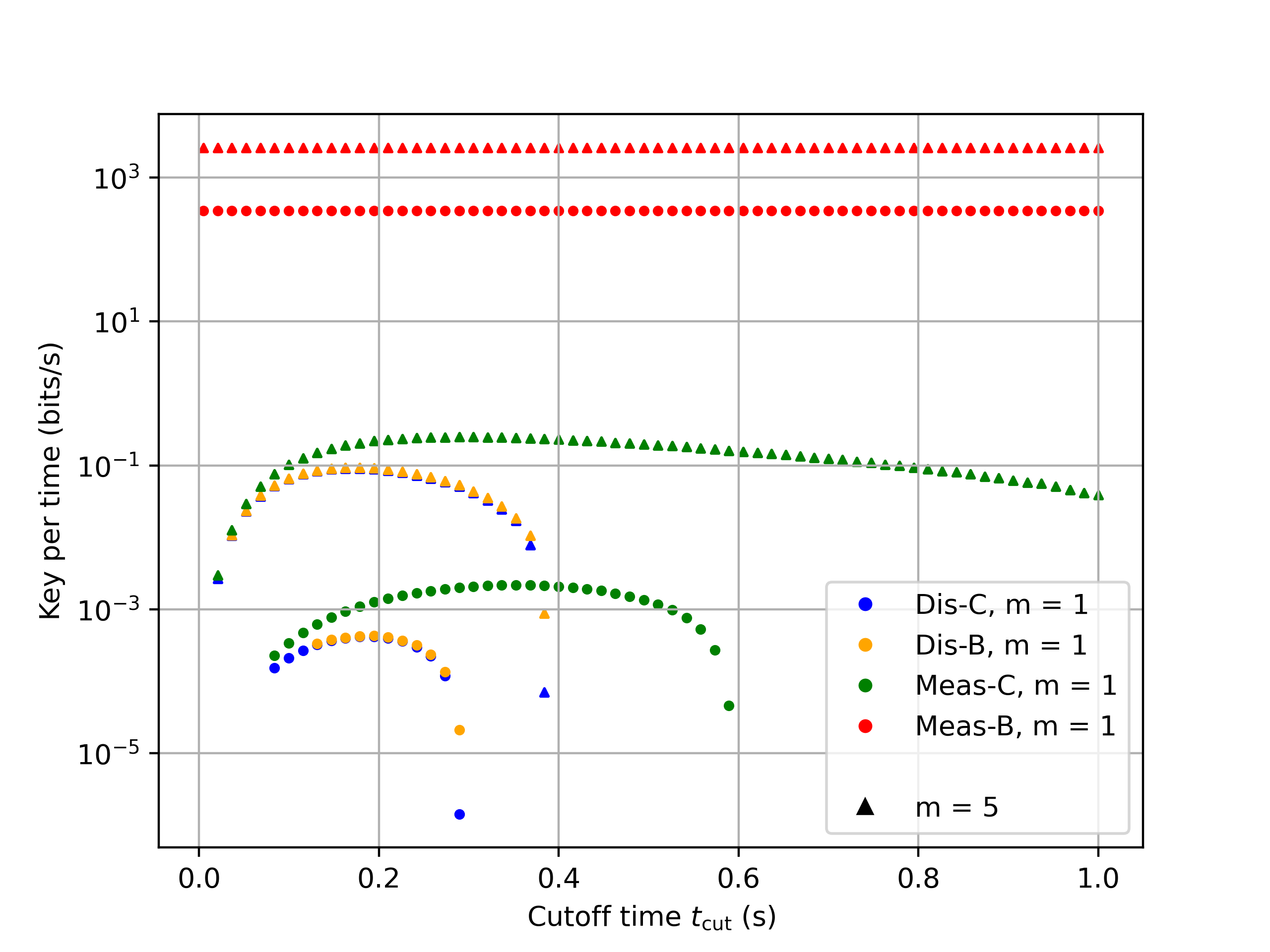}
    \caption{Key rates in dependence of different cutoff times for a completely symmetric star network, when every parties possesses a single memory (round markers) or $m=5$ memories (triangle markers). Other parameters are chosen as $N = 4$, $l = 150\ \mathrm{km}$, $P_{\text{LINK}} = 1$, $F_{\text{INIT}} = 0.99$, $T_P = 10^{-6}\ \mathrm{s}$, $P_D = 10^{-6}$, $T_\text{dp} = 1\ \mathrm{s}$.}
    \label{fig:opt_cut_15}
\end{figure}

Figure~\ref{fig:opt_cut_15} shows the achievable key rates in dependence of the applied cutoff times for a fixed distance of $l = 150\ \mathrm{km}$, considering both the case where each client possesses a single memory and the case where each client has five memories. As expected, the achievable key rates are in general higher when each client is equipped with five memories compared to the case with only one memory. 

The Meas-B scenario can be treated separately, as for the given parameter set we observe no significant change in key rate that could be used to optimized the cutoff time. This behavior stems from the fact that the storage times in this case are significantly shorter than for all other scenarios, and the protocol can operate at a higher rate. The achievable key rate essentially immediately "jumps" to some value, and does not change when increasing the cutoff time. The qubits are never stored long enough in the memories to reach the cutoff time and be discarded.
For all other scenarios, we observe that there exists an optimal cutoff time. We also observe that in the case of multiple memories, cutoff times below $0.08\ \mathrm{s}$ already yield positive key rates, whereas in the single-memory case no key can be generated with the same parameters. This observation is not surprising, as qubits generally spend less time in memory when multiple memories are available, since there are more opportunities to merge them to GHZ state.

\begin{figure}[H]
    \centering
    \begin{minipage}{0.48\textwidth}
        \centering
        \includegraphics[width=\linewidth]{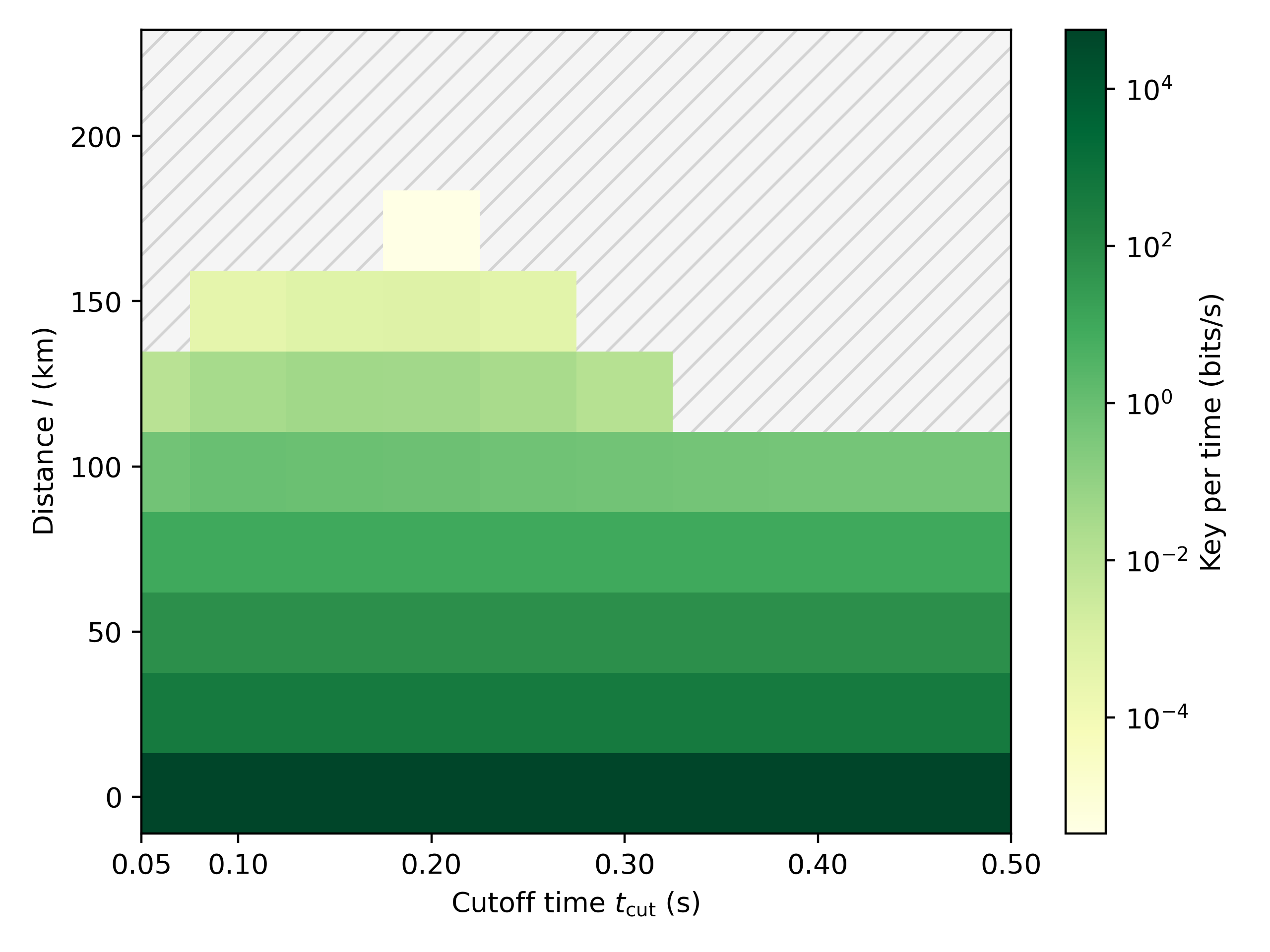}        
        \small (a)
        \label{fig:TCUT_d_dis-C}
    \end{minipage}
    \hfill
    \begin{minipage}{0.48\textwidth}
        \centering
        \includegraphics[width=\linewidth]{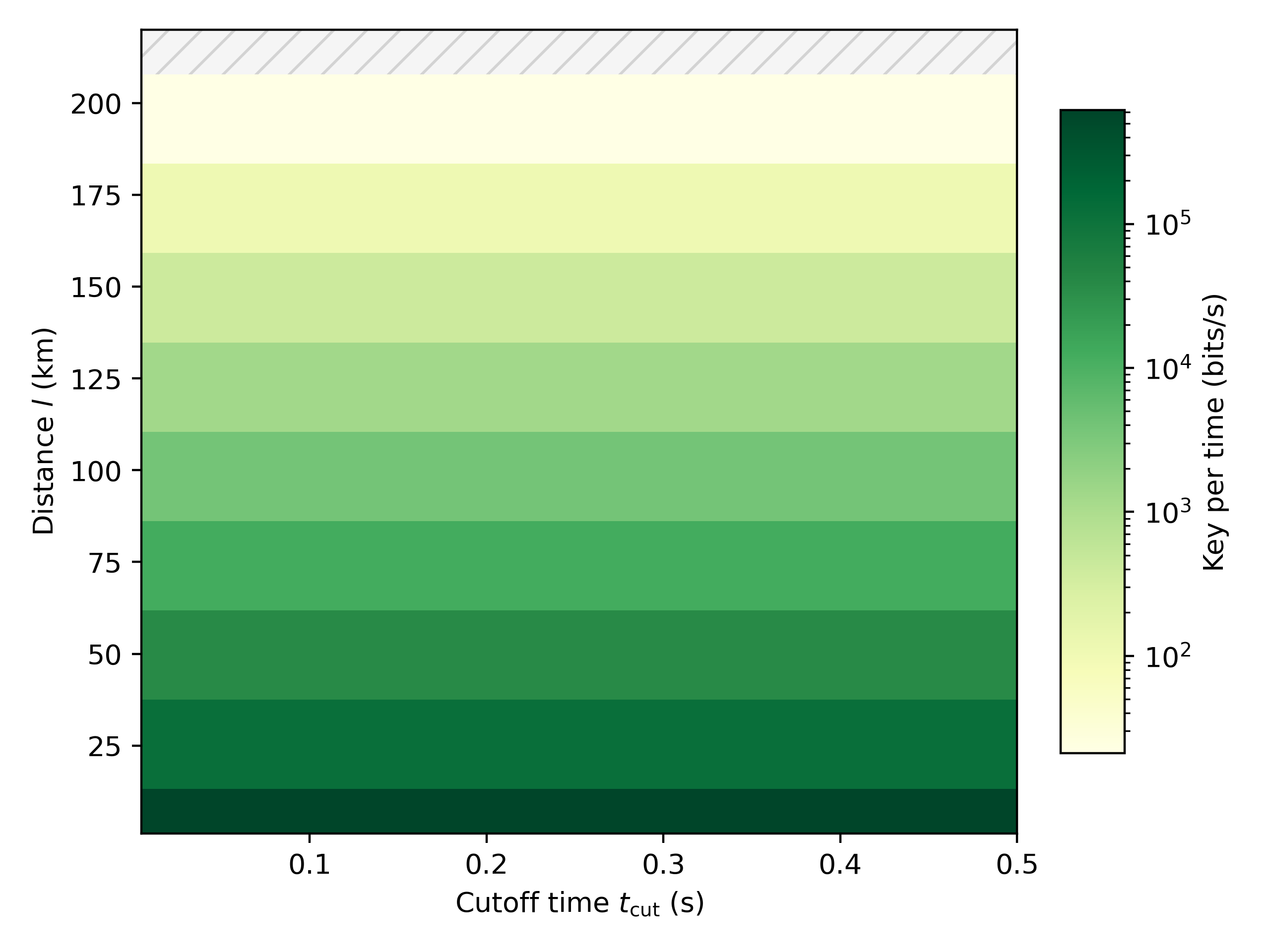}
        \small (b)
        \label{fig:TCUT_d_meas-C.png}
    \end{minipage}
    \caption{Key rates in dependence of the applied cutoff time $t_{\text{\text{cut}}}$ and the distance for the scenario Dis-C (a) and Meas-B (b). Other parameters are chosen as: $N = 4$, $P_{\text{LINK}} = 1$, $F_{\text{INIT}} = 0.99$, $T_P = 10^{-6}\ \mathrm{s}$, $P_D = 10^{-6}$, $T_\text{dp} = 1\ \mathrm{s}$, $m=1$.}
    \label{fig:2D_TCUT_d}
\end{figure}

Note here that in Figure~\ref{fig:opt_cut_15} the distance is fixed to a constant value of $l = 150\ \mathrm{km}$. In Figure~\ref{fig:2D_TCUT_d} we show the key rate in dependence of both the applied cutoff time and the distance. To avoid redundancy in the figures, we select one representative case for each strategy: Dis-C (a) for the Dis approach and Meas-B (b) for the Meas approach.

As can be suspected from Figure~\ref{fig:2D_TCUT_d}, in the Meas-B scenario the achievable secret key rate is independent of the applied cutoff time and depends solely on the distance of a single link. In contrast, in the Dis-C scenario the secret key rate remains independent of the cutoff time only up to a certain distance threshold, beyond which the memory-induced QBER approaches the QBER limit, which ultimately prevents any key distillation. In this regime, the introduction of a cutoff time becomes necessary to enable the CKA protocol, and its value must be chosen to be within a specific range to enable the secure implementation of CKA.

\subsubsection{Varying the number of parties}
\begin{figure}[H]
    \centering
    \begin{minipage}{0.48\textwidth}
        \centering
        \includegraphics[width=\linewidth]{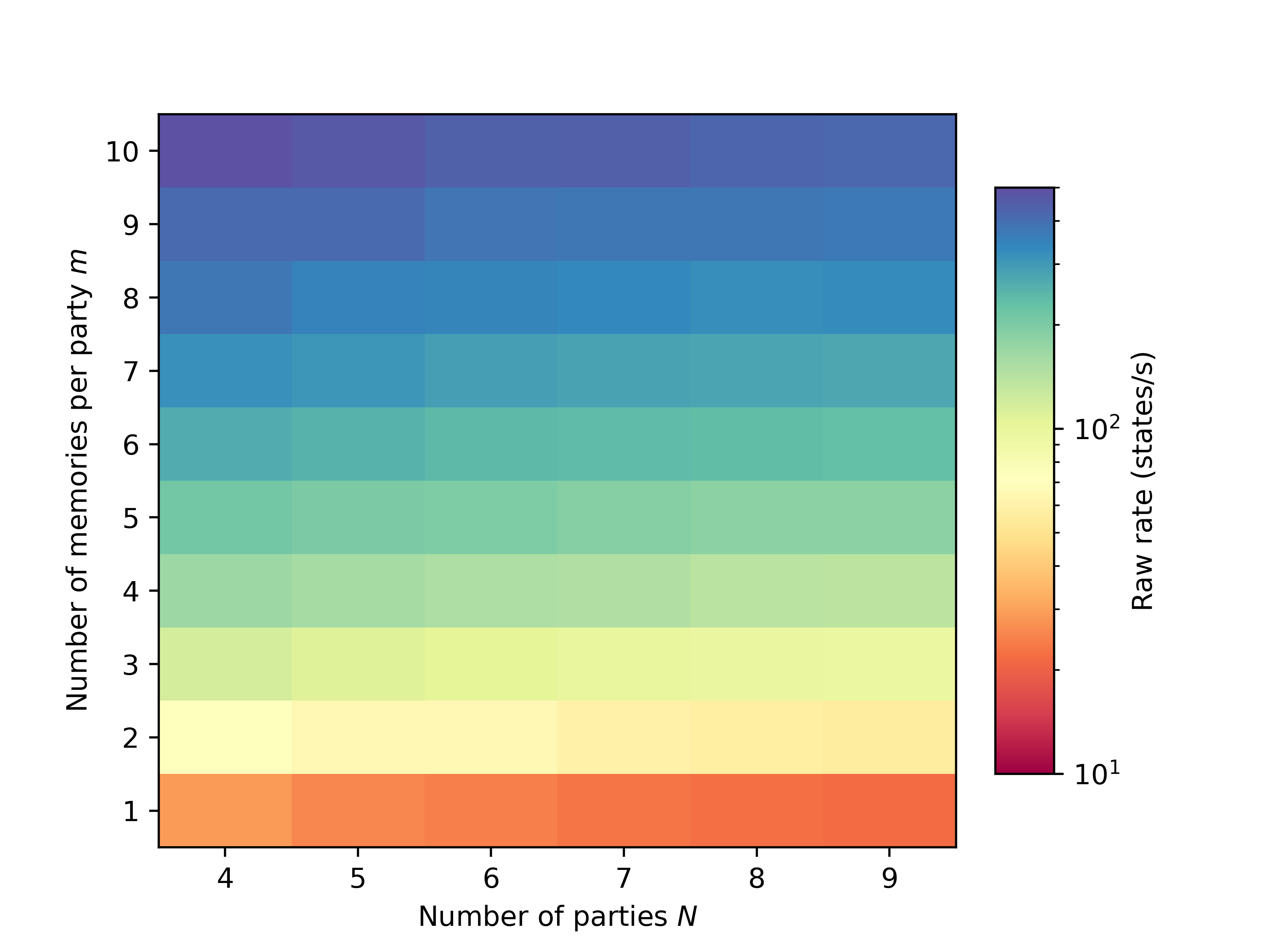}        
        \small (a)
        \label{fig:sym_N_m_raw_rate}
    \end{minipage}
    \hfill
    \begin{minipage}{0.48\textwidth}
        \centering
        \includegraphics[width=\linewidth]{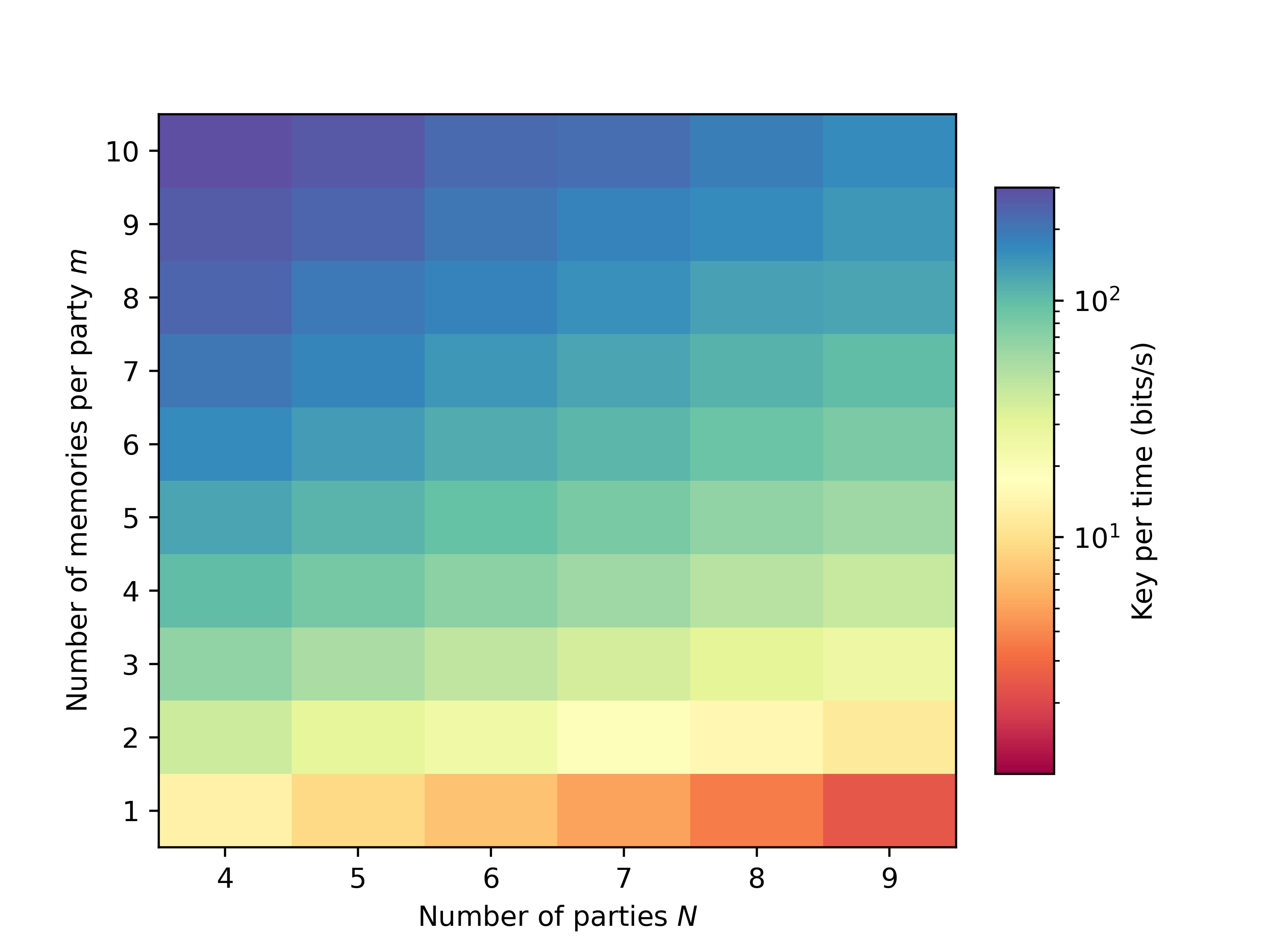}
        \small (b)
        \label{fig:sym_N_m_key}
    \end{minipage}
    \caption{Raw rate (a) and key rates (b) in dependence of number of participants and number of memories for a completely symmetric star network. Other parameters are chosen as: $l = 70\ \mathrm{km}$, $P_{\text{LINK}} = 1$, $F_{\text{INIT}} = 0.99$, $T_P = 10^{-6}\ \mathrm{s}$, $P_D = 10^{-6}$, $T_\text{dp} = 1\ \mathrm{s}$. Note here that this Figure was obtained based on $10^{3}$ GHZ states, as larger $N$ require long run times.
    }
    \label{fig:sym_N_m}
\end{figure}

With an increasing number of parties, the probability of successfully distributing entangled links among all parties decreases. Without memory multiplexing, the raw rate of entanglement distribution (entangled links per attempt) follows a $\left( \ln N\right)^{-1}$ behavior, i.e., it drops fast for small network instances with only a few parties, and it drops slowly with increasing $N$ \cite{kunzelmann_2025}. This can be seen in Figure~\ref{fig:sym_N_m_raw_rate} for different $N$ and fixed $m=1$. With increasing $N$, the decrease in the raw rate becomes less significant, while for a change from $N=4$ to $N=5$ the drop in the raw rate is greater. 

Similar behaviors can be seen for the key rate in Figure~\ref{fig:sym_N_m_key} as the key rate decreases along with the raw rate: 
qubits generally have to be stored for longer periods until each party holds at least one qubit at the center. Consequently, the fidelities are lower, and thus the quantum bit error rates increase, resulting in lower key rates. 
For single memories, a key rate can be distilled for all parties for the chosen parameters, however with a notable decrease in strength. Note that no cutoff time is considered here. 
To counteract this behavior, we introduce memory multiplexing, as this generally increases the fidelity as well as the key rate for a fixed network size (i.e., a fixed number of parties $N$).

\subsection{Asymmetric star network}
\label{sec:results_asymmetric}

Next, we turn to an asymmetric setting, in which the dealer $B_1$ is placed significantly farther from the central station than the other parties; i.e., in Figure~\ref{fig:star-network-topology}, one link is significantly longer than all others. We denote the corresponding distance as $l_{B_1}$. The distance of the $N-1$ shorter links will be denoted as $l_B$. This configuration naturally arises when connecting two distant cities that then branch out to multiple stations within a metropolitan area. Previous work has analyzed this setting analytically~\cite{memmen2023}, but without considering cutoff times for the quantum memories. Note here that the dealer could also be placed at one of the shorter links. This would not change the QBERs in Equation~\ref{CKA_rate}, but the yield, as the long link would only need to be used once. 

\subsubsection{Single memory per client}
\begin{figure}[H]
    \centering
    \subfigure[]{
        \includegraphics[width=0.47\textwidth]{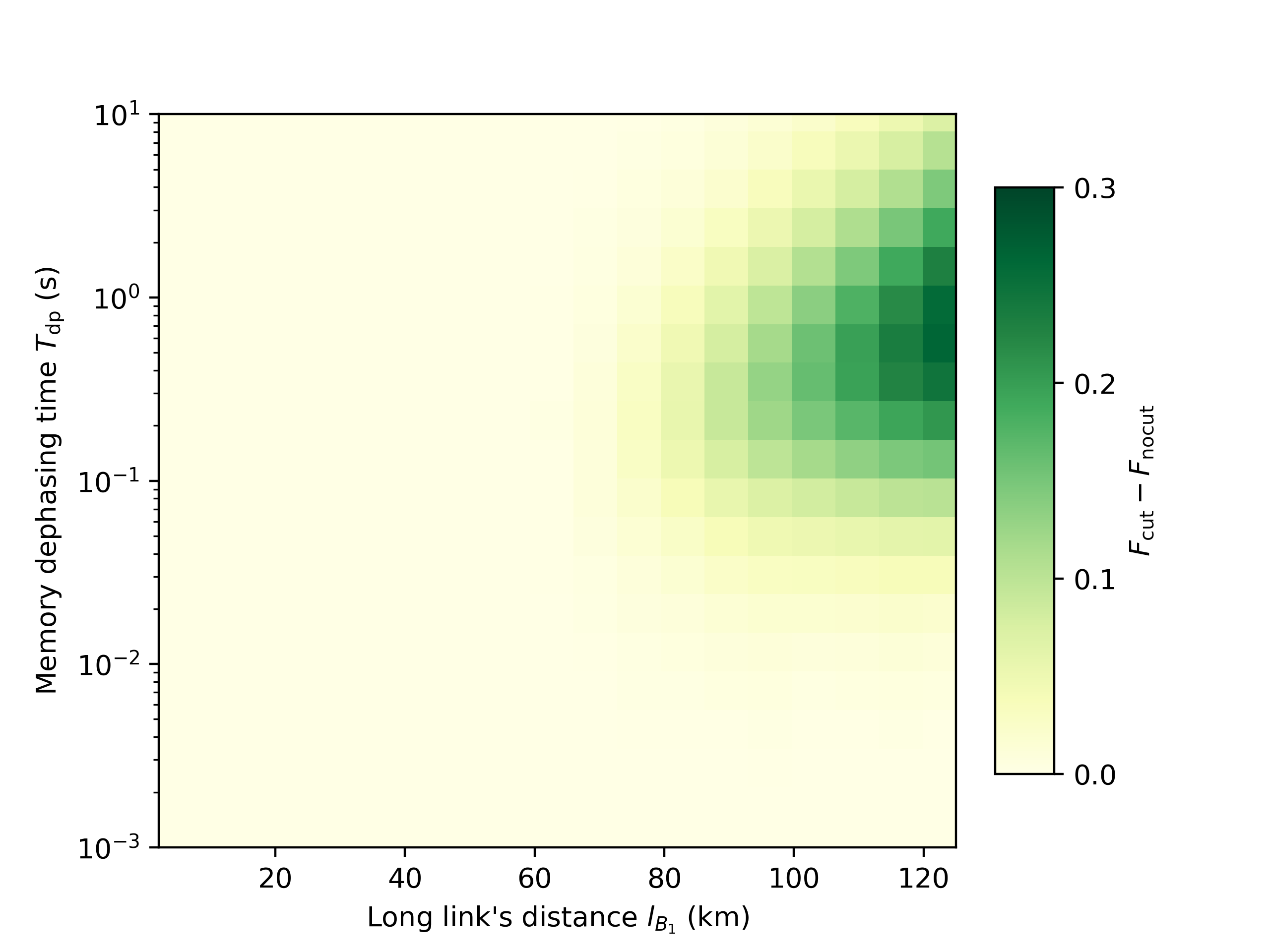}
        \label{fig:asym_delta_fid}
    }
    \hfill
    \subfigure[]{
        \includegraphics[width=0.47\textwidth]{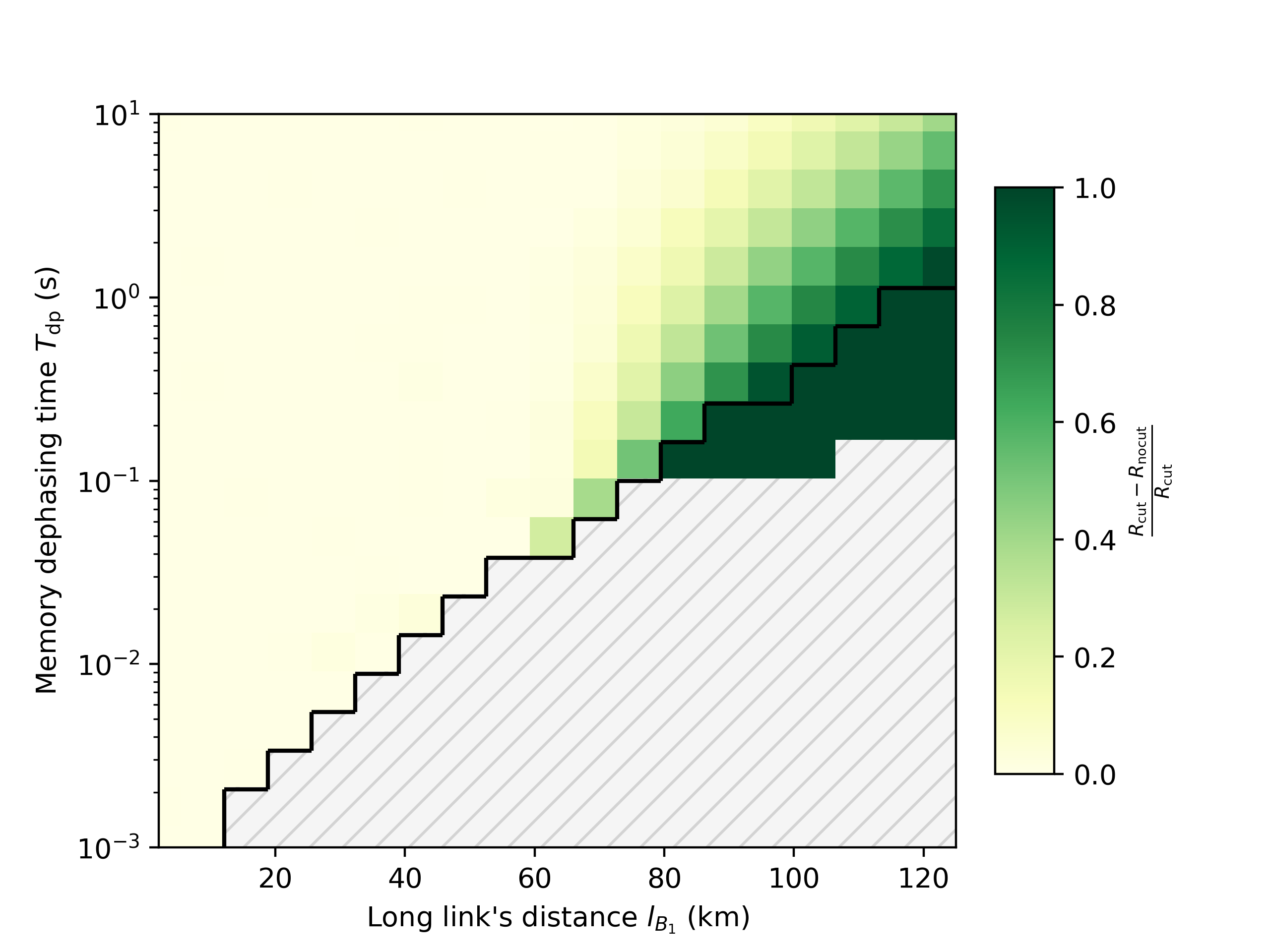}
        \label{fig:asym_key_cut}
    }
        \caption{(a) shows the advantage in terms of fidelity when using a cutoff time of $t_{\mathrm{\text{cut}}}=0.03$ s, compared to the case without a cutoff, as a function of the internal dephasing time $T_\text{dp}$ and the long link's distance $l_{B_1}$ in the asymmetric bottleneck network. (b) shows the corresponding relative advantage in terms of secret key rate when a cutoff time is applied with the same parameter settings. The black contour line indicates the range of parameter settings for which positive key rates remain achievable without the use of a cutoff time. Other parameters are chosen as follows: $N = 4$, $l_{B} = 4\ \mathrm{km}$, $P_{\text{LINK}} = 1$, $F_{\text{INIT}} = 0.99$, $T_P = 10^{-6}\ \mathrm{s}$, $P_D = 10^{-6}$, $T_\text{dp} = 1\ \mathrm{s}$, and $m=1$.}
    \label{fig:asym_cut03}
\end{figure}

We first investigate the scenario in which every client possesses a single memory. For this, we focus on the Dis-C scenario to assess the potential benefits of introducing cutoff times, as this scenario is most susceptible to memory dephasing. In Figure~\ref{fig:asym_delta_fid}, we show the improvement in the fidelity of the target state when applying a cutoff time of $t_{\mathrm{\text{cut}}} = 0.05\ \mathrm{s}$, compared to the case without cutoff times. The improvement is shown as a function of both the long link's distance and the internal memory decoherence time. Figure \ref{fig:asym_delta_fid} shows that cutoff times provide little to no improvement in fidelity for relatively short distances of $l_{B_1}$, since the qubits are not stored long enough for the cutoff to have an effect. For large distances and medium to high memory quality, a significant advantage can be observed. In contrast, for poor memory quality ($T_\text{dp} < 10^{-1}\ \mathrm{s}$), the chosen cutoff time $t_{\text{\text{cut}}}=0.05\ \mathrm{s}$ offers no improvement, as the qubits have already significantly decohered before the cutoff is applied.

Second, we investigate the potential improvement in the achievable secret key rate (see Figure \ref{fig:asym_key_cut}). We plot the improvement obtained with cutoff times, normalized by the achievable secret key rate with cutoff, as a function of the long-link distance and the memory decoherence time. In line with the observed improvement in the fidelity of the target state, we again see that while there is no advantage at short distances, for $l_{B_1} > 80$ km cutoff times enable secure key generation for memory qualities that were insufficient in the case without cutoff.

\subsubsection{Specific scenario connecting universities}
In this section, we consider a representative example of CKA between four universities: Siegen, Duisburg, Wuppertal, and Cologne. The university of Düsseldorf serves as the central station. This configuration effectively corresponds to an asymmetric star network, with one longer link between Siegen and the central node in Düsseldorf (approximately $76\ \mathrm{km}$), and three shorter links connecting to the central node to Cologne ($\sim 31\ \mathrm{km}$), Wuppertal ($\sim 25\ \mathrm{km}$), and Duisburg ($\sim 27\ \mathrm{km}$). Note here that this network is a fictitious example for illustrative purposes and does not correspond to a real-world implementation.

\begin{figure}[H]
    \centering
    \includegraphics[width=0.6\textwidth]{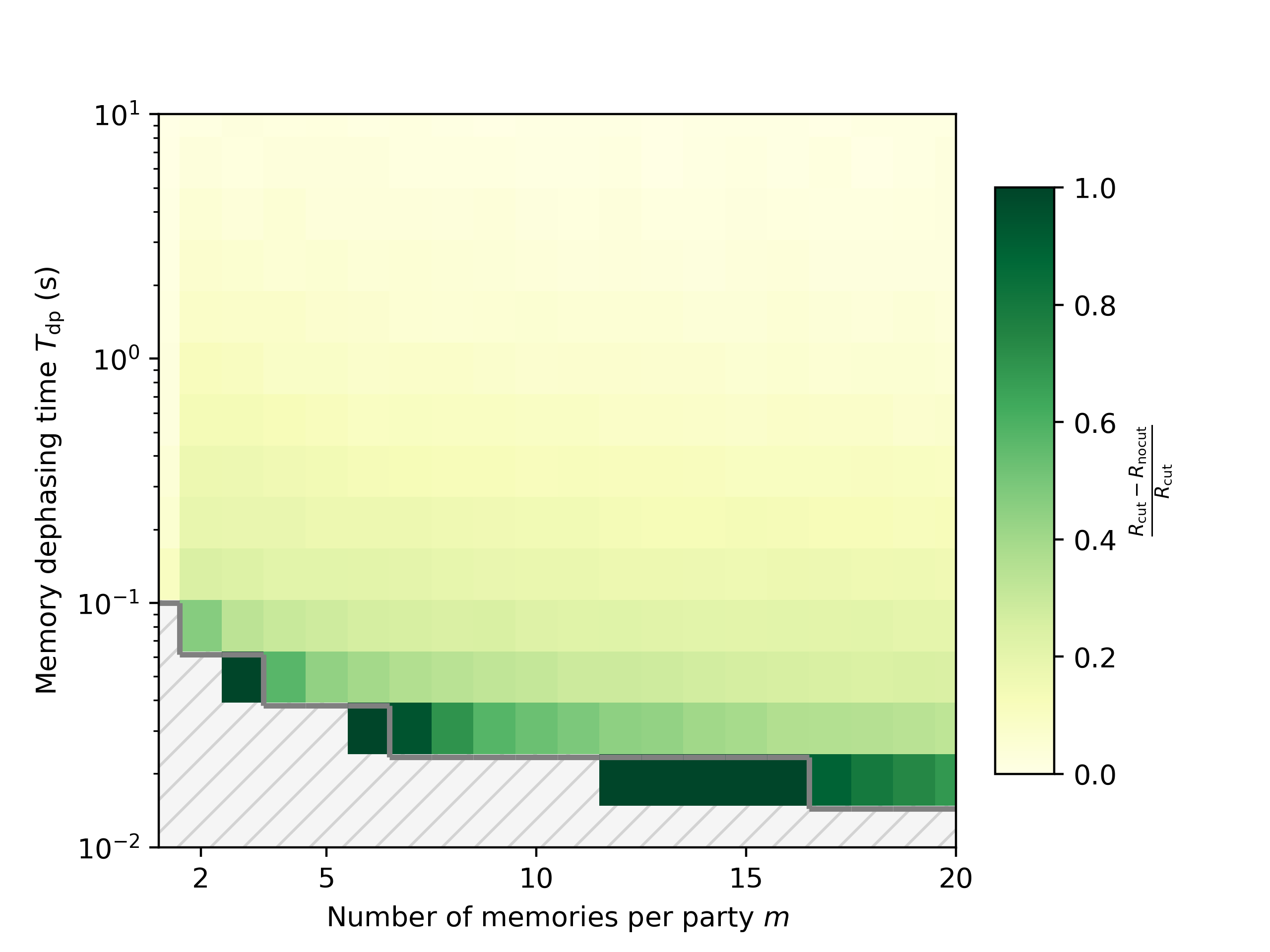}
    \caption{Relative advantage in terms of key rates for the given scenario in dependence of different memory decoherence times $T_\text{dp}$ and memory number $m$. The gray contour line indicates the threshold, where for the case without cutoff time no positive key distillation is possible. The gray striped area indicates the parameter region where no positive key can be distilled. Other parameters are chosen as: $t_{\text{\text{cut}}} = 0.1\ \mathrm{s}$, $P_{\text{LINK}} = 1$, $F_{\text{INIT}} = 0.99$, $T_P = 10^{-6}\ \mathrm{s}$, $P_D = 10^{-6}$, $T_\text{dp} = 1\ \mathrm{s}$.}
    \label{fig:Key_unis_T2_m.png}
\end{figure}

For reasons of clarity, we focus on a single scenario, Dis–C, and compare its performance with and without the application of a cutoff time. Figure~\ref{fig:Key_unis_T2_m.png} shows the effect of incorporating memory multiplexing $m$, as well as the influence of memory quality $T_\text{dp}$, on the overall performance in terms of the relative advantage in the secret key rate for both scenarios.

The results indicate that memory multiplexing can partially mitigate the impact of poor memory quality, as increasing the number of memories increases the achievable key rate. However, below a threshold memory quality of $T_\text{dp} \approx 1.8 \times 10^{-2}\ \mathrm{s}$, even increasing the number of memories per party to up to 20 does not permit secure key generation. For the given range of memories per party ($1 \leq m \leq 20$), this threshold value of $T_\text{dp}$ remains identical in both cases, with and without a cutoff time. Nonetheless, when a cutoff time is introduced, a smaller number of memories already suffices to compensate for poor memory quality.

\section{Summary and outlook}
\label{sec:summary}

In this work, we have comprehensively analyzed the generation of GHZ states and the corresponding achievable secret key rates of a CKA implementation in both symmetric and asymmetric star-network topologies. Our results cover a wide range of practically relevant scenarios, including different state-distribution strategies and placements of bipartite entanglement sources, the use of memory multiplexing, and the introduction of cutoff times for the deployed quantum memories.
The simulation framework used for this purpose, ReQuSim, has proven to be an extremely versatile tool and enables the exploration of architectures and strategies that are difficult to evaluate analytically.

Our results highlight several key insights. First, cutoff times play a central role in long-distance quantum networks, but they must be carefully optimized for the specific architecture and deployed strategy. Second, even small variations in the network configuration can lead to markedly different results. Finally, memory multiplexing can provide a valuable performance advantage, and can, to some extent, compensate for imperfections such as limited memory coherence or an increasing number of network participants. These results are especially consequential for the case where the nodes do not possess a photon source, but only detectors. Such a case is very attractive from a practical perspective but our results indicate that strategy optimization is of the utmost importance in this setting.
Overall, our findings illustrate the importance of comprehensive simulation tools for guiding the development of future quantum network architectures.

With regard to the star networks considered in this work, there are a few potential directions for further investigation. Further protocol optimizations, such as partially merging a subset of Bell pairs to a GHZ state rather than waiting until Bell pairs with all parties have been established, could reduce the number of qubits that are affected by memory noise. This could be of particular interest for those hardware platforms for which the merging operation itself is probabilistic. Furthermore, our approach could be interfaced with other numerical tools that allow for more sophisticated optimization or to consider other figures of merit, e.g., finite key rate analysis. Yet another direction would be to consider more involved noise models arising from specific physical setups.


\section*{Code and data availability}
The simulation is based on ReQuSim, an open source Python package available from the Python Package Index and at Ref.\ \cite{requsim_zenodo}. The code we used for the scenarios in this work is archived at DOI  \href{https://doi.org/10.5281/zenodo.20085326}{10.5281/zenodo.20085326}. The raw data from the simulation is available upon reasonable request.

\section*{Acknowledgements}
J.~M. acknowledges funding via the Q-net-Q Project (supported by BMFTR and the EU's Digital Europe Programme No.~101091732), the Quant-GPlCz Project (supported by BMDV grant No.~19OS25001A and the EU's Horizon Europe Programme No.~101249338), and Berlin Quantum. J.~W., N.~W. and J.~E. acknowledge support from the BMFTR via the projects QR.X and QR.N (for which this is the result of a joint node collaboration), the Cluster of Excellence ML4Q (for which this is again the result of a joint node collaboration),
Berlin Quantum
(for which this is once more the result of a joint node collaboration)
and the European Research Council (DebuQC). J.~W. acknowledges support from the Austrian Science Fund (FWF). This research was funded in whole or in part by the Austrian Science Fund (FWF) 10.55776/P36009 and 10.55776/P36010. For open access purposes, the author has applied a CC BY public copyright license to any author-accepted manuscript version arising from this submission.

\bibliographystyle{abbrv}
\bibliography{literature}

\appendix
\section{Measurement procedure}

In the protocol it is mentioned that the central station $C$ connects the $N$ Bell pairs it shares with the clients to a GHZ state.
This can be done by performing an $n$-qubit measurement in the basis
\begin{equation}
\ket{\Psi_{j\mathbf{i}}} = \frac{1}{\sqrt{2}} \left(\ket{0,\mathbf{i}} + (-1)^j \ket{1, \bar{\mathbf{i}}} \right)
\end{equation}
with $\mathbf{i} =i_2i_3\dots i_N$ being a bit string of length $(n-1)$  and $\bar{\mathbf{i}}$ denoting its inversion.
After obtaining a measurement outcome that corresponds to $\Psi_{j\mathbf{i}}$, the following correction operation needs to be applied by the clients $B_k$ in order to recover the standard $n$-party GHZ state 
\begin{equation}
Z^j_{B_1} \prod_{k=2}^N X^{i_k}_{B_k}
\end{equation}
For the purposes of our numerical simulation we use a different but equivalent way to calculate the resulting density matrix. In order to avoid calculating a $2^{2N} \times 2^{2N}$ density matrix, we instead successively merge each individual Bell pair to a growing GHZ state, via a POVM with measurement operators $\ket{0}\bra{0,0} + \ket{1}\bra{1,1}$ and $\ket{0}\bra{0,1} + \ket{1}\bra{1,0}$ (with the outcome corresponding to the latter requiring a Pauli correction). Finally, an $X$-measurement is performed on the final remaining qubit at $C$.

\section{Comparison to bipartite schemes}
In a bottleneck network, CKA can also be realized solely through bipartite entanglement \cite{Epping_2017, memmen2023}. In this setting, a designated player (the dealer) establishes bipartite QKD links with all other parties and subsequently uses these links to encode the conference key. The use of multipartite entanglement has been shown to offer a performance advantage in that comparison, as it exploits the network topology more efficiently and requires fewer network uses than the bipartite approach (1 compared to $N - 1$). This advantage, however, is limited to regimes of low loss and low noise, since the multipartite protocol is more sensitive to noise and necessitates the simultaneous transmission of all qubits. 

\begin{figure}[H]
    \centering
    \includegraphics[width=0.6\textwidth]{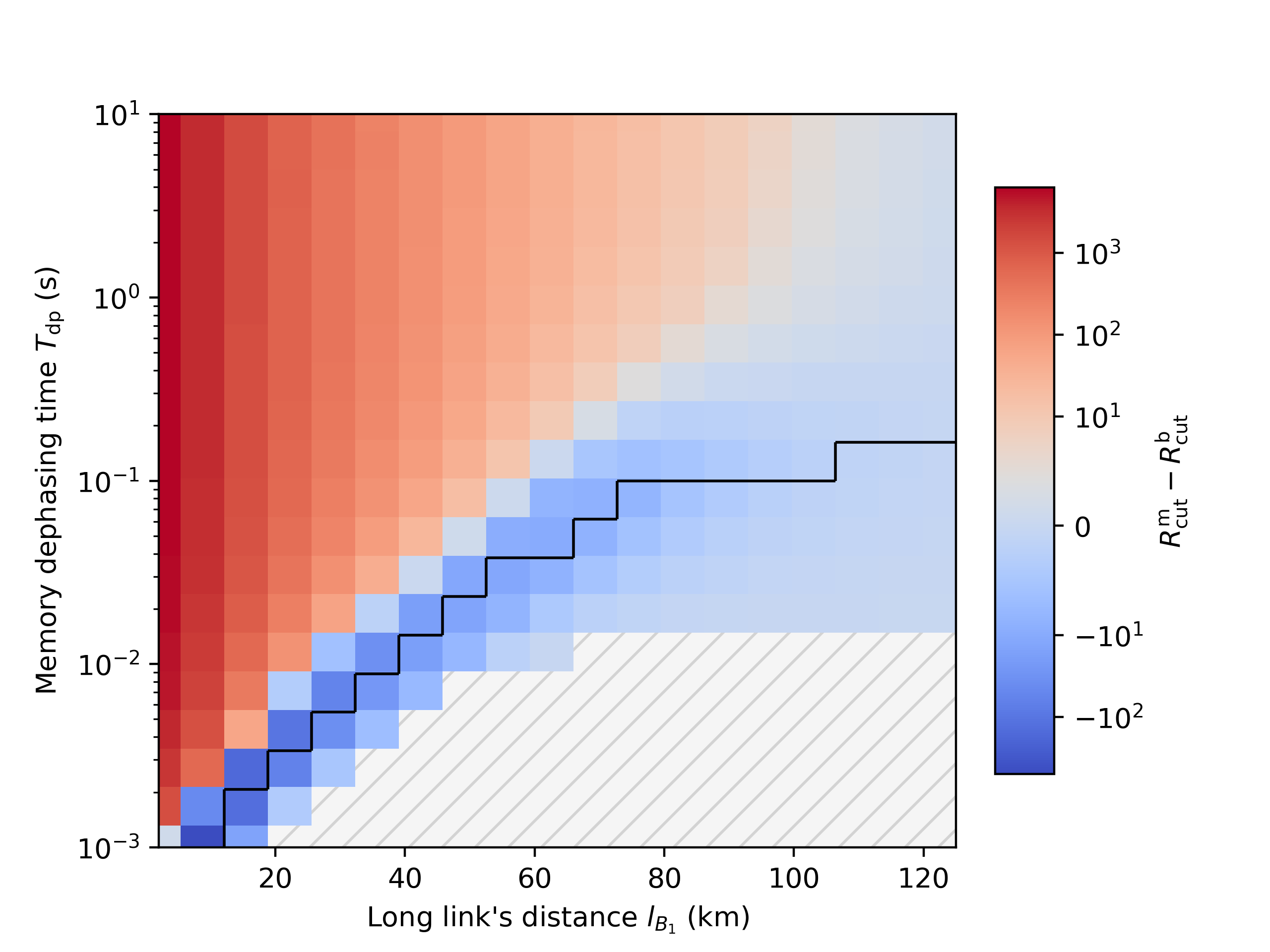}
    \caption{Absolute advantage in terms of key rates of a multipartite protocol with $N = 4$ parties compared to a bipartite implementation when both make use of a cutoff time in dependence of different memory decoherence times $T_\text{dp}$ and distances of the longest link $l_{B_1}$. Red areas correspond to a multipartite advantage, blue areas to a bipartite advantage. The black line indicates the threshold beyond which no positive key can be distilled in the multipartite case, the gray area indicates that no positive key can be distilled in either protocol, bipartite and multipartite. Other parameters are chosen as: $l_B = 4\ \mathrm{km}$, $t^m_{\text{\text{cut}}} = 0.05$s, $t^b_{\text{\text{cut}}} = 0.015\ \mathrm{s}$, $P_{\text{LINK}} = 1$, $F_{\text{INIT}} = 0.99$, $T_P = 10^{-6}\ \mathrm{s}$, $P_D = 10^{-6}$, $T_\text{dp} = 1\ \mathrm{s}$.}
    \label{fig:asym_bibench_with_cut}
\end{figure}

Recent work has further demonstrated that this multipartite advantage can be enhanced in asymmetric star networks when quantum memories are available \cite{memmen2023}. In the following, we investigate how the multipartite advantage behaves in an asymmetric network with quantum memories when these memories are equipped with a finite cutoff time. Figure~\ref{fig:asym_bibench_with_cut} shows the absolute advantage in terms of secret key rate for a CKA protocol based on multipartite entanglement compared to one relying on bipartite entanglement for $N=4$, in dependence of the long link's distance and the memory dephasing time $T_\text{dp}$. In both settings, a cutoff time is applied to the quantum memories. However, the cutoff time applied in the bipartite setting should be shorter than that in the multipartite protocol, as qubits in the bipartite case spend less time in memory due to the fact that only one other link needs to be established. The data for the bipartite setting was obtained by running the simulation for $N=2$ and then incorporating the network uses by dividing the resulting key rate by $(4-1)=3$. 

As expected, the multipartite advantage is more prominent for fairly good memory qualities. This occurs because the QBERs are more affected by poor memory quality in the multipartite case, as in that case they depend on multiple qubits. Nonetheless, with the cutoff times applied, a multipartite advantage can still be observed for memory coherence times of $T_\text{dp} = 1\ \mathrm{s}$ for distances up to $l_{B_1} \approx 80\ \mathrm{km}$.

\end{document}